\documentclass[twocolumn,preprintnumbers,amsmath,amssymb,pra]{revtex4}

\usepackage{graphicx,amsmath}
\usepackage{dcolumn}
\usepackage{bm}
\usepackage{amssymb}
\usepackage{epstopdf}
\usepackage{color}
\begin{document}

\title{Stimulated emission for a three-level artificial atom in waveguide quantum electrodynamics}

\begin{abstract}

This work is devoted to a theoretical study of the interaction of a quantum three-level ladder system with a continuous electromagnetic field in a one-dimensional open waveguide. We consider a situation closely resembling stimulated emission — the scattering of a single-photon exponential pulse by an excited three-level emitter. Using the real-space formalism, we obtain an analytical expression for the wavefunction of our system. We show that, for the low anharmonicity of a three-level system inherent to a transmon (the most common type of artificial atom), the influence of the third level can have a significant effect on the system's behavior. Namely, the presence of the third level largely suppresses the stimulated emission effect if the relative anharmonicity is lower than $| \alpha_r | < 4\%$. However, for a three-level system, it is possible to obtain correlation effects such as photon bunching in the reflected field, which can also be 'switched off' by tuning the incident photon parameters.

\end{abstract}

%\pacs{84.40.Az,~ 84.40.Dc,~ 85.25.Hv,~ 42.50.Dv,~42.50.Pq}
\date{October 7, 2026}

\author{O. A. Chuikin}\email{ChuikinOA@yandex.ru}
\affiliation{Department of Applied and Theoretical Physics, Novosibirsk State Technical University, Novosibirsk 630073, Russia}

\author{Ya. S. Greenberg}
\affiliation{Department of Applied and Theoretical Physics, Novosibirsk State Technical University, Novosibirsk 630073, Russia}

\author{V. V. Taravkova}
\affiliation{Department of Applied and Theoretical Physics, Novosibirsk State Technical University, Novosibirsk 630073, Russia}

\author{O. V. Kibis}
\affiliation{Department of Applied and Theoretical Physics, Novosibirsk State Technical University, Novosibirsk 630073, Russia}

\maketitle

\section{Introduction}

Quantum three-level systems play an important role in quantum optics and photonics \cite{Gu17}, and they are also of particular interest in the emerging field of waveguide quantum electrodynamics \cite{Sheremet23}. In contrast with cavity QED, which usually focuses on resonant interaction of photons with natural atoms \cite{Raimond01}, nanostructures \cite{Savenko12}, or condensed-matter materials such as graphene \cite{Kibis11}, waveguide QED describes broadband interaction of atoms or nanostructures with electromagnetic continuum fields \cite{Roy17}.

The interaction of three-level emitters (3LEs) with photon fields can induce a number of interesting effects, such as electromagnetically induced transparency \cite{Abdumalikov10}, Autler-Townes splitting \cite{Sillanpaa09}, two-photon resonance fluorescence under single-tone \cite{Gasparinetti19} or bichromatic pumping \cite{Gunin25}, the giant cross-Kerr effect \cite{Hoi13a}, and many others. Moreover, 3LEs can be utilized in more practical scenarios in modern quantum technologies, such as microwave photon routing \cite{Hoi11}, on-chip quantum amplification \cite{Astafiev10b}, or the generation of correlated photon pairs \cite{Sathyamoorthy16}.

There is a vast number of theoretical works dedicated to the interaction of three-level systems with continuous fields in one-dimensional (1D) waveguides. Polarization control of a single photon in a waveguide via its interaction with a chain of three-level $\Lambda$-type atoms was described in \cite{Tsoi09}. Single-photon scattering from driven and undriven $\Lambda$-type or V-type systems was studied in \cite{Witthaut10} for the stationary regime and linear dispersion, and in \cite{Martens13} for both stationary and time-dependent regimes in the case of nonlinear dispersion. In \cite{Roy11}, it was shown that two photons scattered from a driven $\Lambda$-type atom exhibit stronger correlations compared to those in a regular two-level system. In a follow-up paper \cite{Roy14}, a more detailed study of electromagnetically induced transparency for a driven $\Lambda$-system and the effects of bunching and antibunching in the transmitted field was performed. Photon-photon correlations and photon entanglement were studied in \cite{Zheng12} for driven $\Lambda$-type three-level and N-type four-level systems, and in \cite{Li15} for the scattering of two distinguishable photons from a ladder-type atom. Photon-state purification due to the interaction of a two-photon wavepacket with a $\Lambda$-type three-level system was studied in \cite{Lopes24}.

As one can see, most research works are focused on $\Lambda$- or V-type systems because they can be readily realized with natural atoms. However, in recent years, another type of quantum emitter has gained popularity, known as an artificial atom — a technologically engineered nanostructure with an energy-level spectrum similar to that of natural atoms, but with much greater possibilities for parameter tuning and control. Artificial atoms are usually implemented using quantum dots \cite{Lodahl15} or superconducting circuits \cite{Blais21}. The most common type of artificial atom today is the superconducting qubit known as the transmon \cite{Koch07}, which has a ladder-type energy-level structure with low anharmonicity. Despite the fact that only the two lowest levels of the transmon are used in most experiments and practical applications, the influence of the third level can be quite strong and often needs to be taken into account \cite{Greenberg26}. Thus, transmons can be effectively used to study the interaction of three-level ladder systems with continuous photon fields in one-dimensional waveguides.

In this work, we study one of the most basic models of stimulated emission — a single-photon pulse with an exponential shape incident on an excited artificial atom inside a 1D open waveguide. From a formal standpoint, it is sufficient to describe the atom in this scenario as a regular two-level system, or a qubit \cite{Rephaeli12}. However, as stated above, the most typical type of artificial atom, i.e., the transmon, has a low-anharmonicity three-level structure, and thus the influence of the third level must be taken into account for a more rigorous theory.

In our work, we use the real-space formalism to obtain an analytical expression for the wavefunction in the case of an arbitrary single-photon pulse incident on an excited 3LE. We then study the specific case of an exponential pulse, which may be of particular interest since this type of photon pulse is usually emitted by atoms in a process of spontaneous emission \cite{Valente12}. Using the obtained analytical expressions, we analyse the scattering dynamics in such a system, showing the difference between three-level and two-level systems. From our findings, we conclude that the influence of the third level can be of great importance; for example, it can suppress stimulated emission or induce bunching in a reflected two-photon state.

The paper is organized as follows. In Section II, a general description of the studied model is given based on the real-space representation of photon fields. In Section III, we find a general solution for the wavefunction of our system for an arbitrary incident pulse. Then, we consider the particular case of an exponential pulse and study the scattering dynamics of atomic excitation and single-photon states in Section IV. Section V is devoted to the suppression of stimulated emission by the third level — an effect that can be observed by analysing two-photon states. Section VI describes correlation effects that appear in two-photon states and result in spatial bunching. We also show that reflected fields with two left-moving photons can exhibit bunching and non-bunching behaviour depending on the characteristics of the initial pulse. Closing thoughts and conclusions are presented in the final Section VII.

\section{Real-space description of the model}

\begin{figure}
\includegraphics[width=0.5\textwidth]{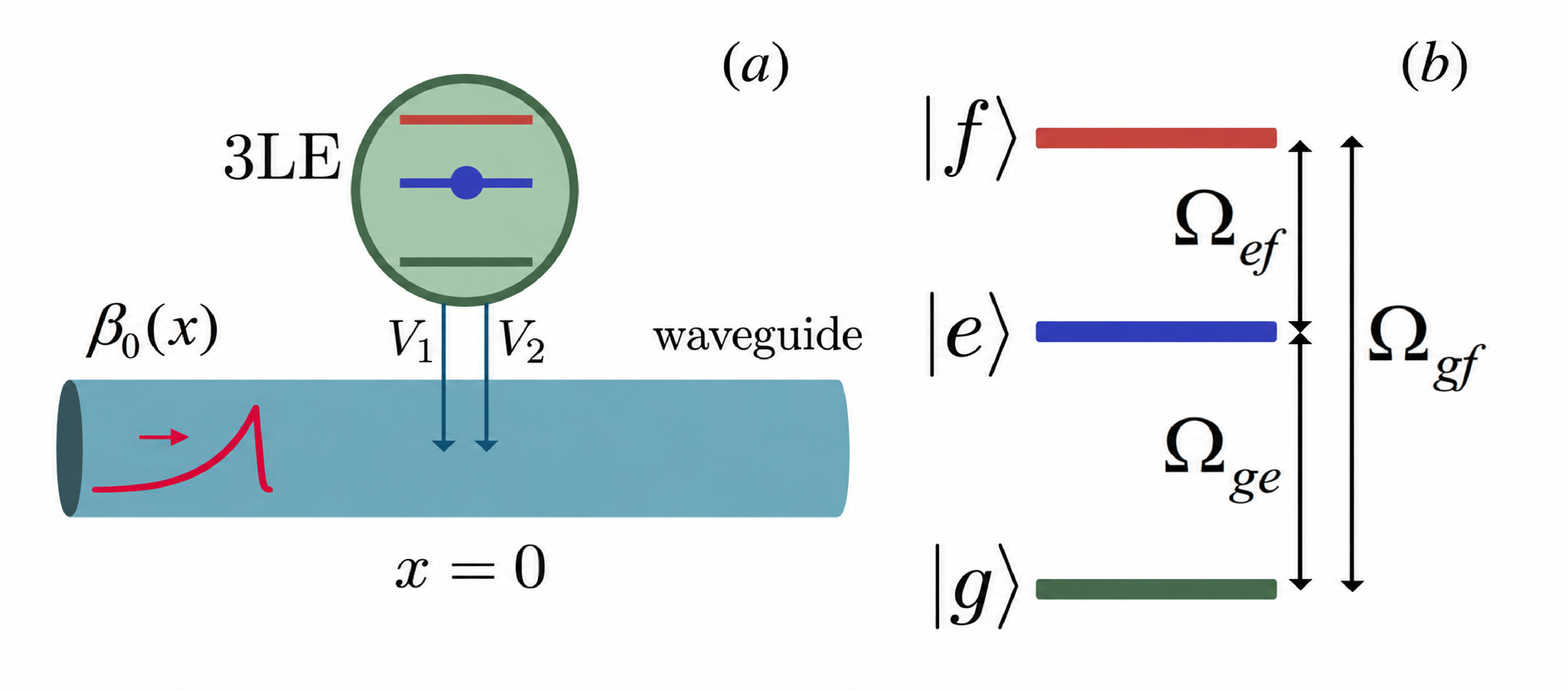}\\
  \caption{ (a) Schematic of the considered system: a three-level emitter (3LE) coupled to the center of a one-dimensional waveguide at point $x=0$. The coupling strengths for the middle and upper levels are $V_1$ and $V_2$, respectively. At the initial moment, a single-photon pulse $\beta_0(x)$ (which we later take to be an exponential pulse) is incident from the left side of the waveguide onto a 3LE excited to the level $\left| e \right\rangle$. (b) Energy-level diagram of the bare 3LE and the corresponding resonant frequencies.}  \label{fig1} 
  \end{figure}

We consider a ladder-type three-level emitter (3LE) located at the point $x=0$ in a one-dimensional waveguide (see figure \ref{fig1}(a)). In the real-space representation \cite{Shen07} and under the rotating wave approximation (RWA), it can be described by the following Hamiltonian (here and throughout the text, we assume that $\hbar = 1$):
  \begin{equation} \label{hamiltonian}
\begin{gathered}
H = \Omega_{ge} \left| e \right\rangle \left\langle e \right| + \Omega_{gf} \left| f \right\rangle \left\langle f \right| 
\\
- i \upsilon_g \int\limits_{-\infty}^{+\infty} dx \; a^\dagger (x) \frac{\partial }{\partial x} a(x) + i \upsilon_g \int\limits_{-\infty}^{+\infty} dx \; b^\dagger (x) \frac{\partial }{\partial x} b(x)
\\
+ \int\limits_{-\infty}^{+\infty} dx \; V_1 \delta (x) \left( a^\dag (x) + b^\dag (x) \right) \left| g \right\rangle  \left\langle e \right| + h.c.
\\
+ \int\limits_{-\infty }^{+\infty} dx \; V_2 \delta (x) \left( { a^\dag (x) + b^\dag (x) } \right) \left| e \right\rangle \left\langle f \right| + h.c.
\end{gathered} 
  \end{equation}
  
Here, the states $\left| g \right\rangle, \left| e \right\rangle$ and $\left| f \right\rangle$ are the ground, first-excited, and second-excited states of the emitter. The corresponding resonant frequencies of the first and second excited states are $\Omega_{ge}$ and $ \Omega_{gf}$, respectively, and $V_1, \; V_2$ are their coupling constants. The energy levels and corresponding frequencies are shown in figure \ref{fig1}(b). $\upsilon_g$ is the group velocity of photons inside the waveguide, and $\delta(x)$ is the Dirac delta function. $a^\dagger(x)$ ($b^\dagger(x)$) are the creation operators in the $x$-representation for right-moving (left-moving) photons that satisfy the standard commutation relation:
  \begin{equation}
\begin{gathered}
\left[ a(x), a^\dagger (x') \right] = \left[ b(x), b^\dagger (x') \right] = \delta(x-x'),
\\
\left[ a(x), b^\dagger (x') \right] = \left[ a(x), b (x') \right] = 0.
\end{gathered} 
  \end{equation}
Hamiltonian (\ref{hamiltonian}) is written in the continuous limit \cite{Blow90}, which is a convenient way to describe photonic fields in open waveguides. However, due to this, the photon operators, which are usually dimensionless for discrete modes, have dimensions of inverse square root of length in the continuous limit, $a(x) \sim 1/\sqrt{x}$. The dimensions of the coupling constants $V_1$ and $V_2$ are correspondingly $V \sim \omega \sqrt{x}$.

In this paper, we address the problem of stimulated emission, in which a single-photon pulse is incident on an excited emitter. Thus, we must work in a two-excitation subspace, and the general wavefunction of our system is given by:
  \begin{equation}
\begin{gathered}
\left| \Psi (t) \right\rangle  = \alpha (t) e^{-i\Omega_{gf}t } \left| f,0 \right\rangle 
\\
 + \int {dx} e^{ -i\Omega_{ge}t} \beta_R(x,t) a^\dagger(x) \left| e,0 \right\rangle 
\\
 + \int {dx} {e^{ - i{\Omega _{ge}}t}}{\beta _L}(x,t){b^\dag }(x)\left| {e,0} \right\rangle 
 \\
 + \iint {dx_1} {dx_2} {\gamma_{R,R}}(x_1,x_2,t) \frac{1}{\sqrt 2} {a^\dag }(x_1){a^\dag }({x_2}) \left| g,0 \right\rangle  
 \\
 + \iint {d{x_1}} {d{x_2}\;} {\gamma _{L,L}}({x_1},{x_2},t)\frac{1}{{\sqrt 2 }}{b^\dag }({x_1}){b^\dag }({x_2})\left| {g,0} \right\rangle 
 \\
 + 2\iint {d{x_1}} {d{x_2}\;} {\gamma _{R,L}}({x_1},{x_2},t)\frac{1}{{\sqrt 2 }}{a^\dag }({x_1}){b^\dag }({x_2})\left| {g,0} \right\rangle ,
\end{gathered} \label{wavefunc}
  \end{equation}
where we assume that the integration is taken over the entire waveguide, ideally from $-\infty$ to $+\infty$. Generally, the wavefunction (\ref{wavefunc}) contains three types of probability amplitudes. The first amplitude, $\alpha(t)$, corresponds to a fully excited 3LE in the state $\left| f \right\rangle$. The other two amplitudes, $\beta_R(x,t)$ and $\beta_L(x,t)$, correspond to an excited atom in the state $\left| e \right\rangle$ and a single photon propagating in the right or left direction. The remaining three amplitudes, $\gamma_{R(L),R(L)}(x_1,x_2,t)$, correspond to two photons moving either to the left or to the right, with the atom in the ground state $\left| g \right\rangle$. The state $\left| 0 \right\rangle$ is the vacuum state of the waveguide. Since we work in the continuous limit of the real-space representation, the photon-related amplitudes have dimensions $\beta(x,t) \sim 1/\sqrt{x}$, and $\gamma(x_1, x_2, t) \sim 1/x$. The coefficient $1/\sqrt 2$ in the last three terms is a normalization constant for a two-photon Fock state.

Since photons are bosonic particles, we assume that they are practically indistinguishable. Moreover, photon indistinguishability plays a major role in the process of stimulated emission \cite{Sun07}. Thus, the following relations can be justified:
   \begin{equation}
\begin{gathered}
\gamma_{i,i}(x_1,x_2,t) = {\gamma_{i,i}}(x_2,x_1,t); \qquad i = L,R
\\
\gamma_{R,L}(x_1,x_2,t) = \gamma_{L,R}(x_2,x_1,t);  \label{indist}
\end{gathered}
   \end{equation}
Due to these relations, the last term in (\ref{wavefunc}) has a factor of 2.   

\section{General solution for the probability amplitudes}

Using the standard time-dependent Schr\"oedinger equation with Hamiltonian (\ref{hamiltonian}) and wavefunction (\ref{wavefunc}), one can obtain the following differential equations for the amplitudes:
   \begin{equation}
\frac{d}{dt} \alpha (t) =  -iV_2 e^{i\Omega_{ef}t} \left({ \beta_R(0,t) + \beta_L(0,t) }\right),   \label{eq_alpha}
   \end{equation}
   \begin{equation}
\begin{gathered}
\frac{d}{dt} \beta_R(x,t) + \upsilon_g \frac{\partial}{\partial x} \beta_R(x,t) =  -iV_2 \delta (x) {e^{-i\Omega_{ef} t } } \alpha (t)
\\
 - i\frac{2}{\sqrt 2} V_1 e^{i\Omega_{ge}t}  \left[{ \gamma_{R,R}(x,0,t) + \gamma_{R,L}(x,0,t) }\right],
\end{gathered}  \label{eq_betaR}
   \end{equation}   
   \begin{equation}
\begin{gathered}
\frac{d}{dt} \beta_L(x,t) - \upsilon_g \frac{\partial}{\partial x} \beta_L(x,t) =  -iV_2 \delta (x) {e^{-i\Omega_{ef} t } } \alpha (t)
\\
 - i\frac{2}{\sqrt 2} V_1 e^{i\Omega_{ge}t}  \left[{ \gamma_{L,L}(x,0,t) + \gamma_{R,L}(0,x,t) }\right],
\end{gathered}   \label{eq_betaL}
   \end{equation}
   \begin{equation}
 \begin{gathered}
\left({ \frac{d}{dt} + \upsilon_g \frac{\partial}{\partial x_1} + \upsilon_g \frac{\partial}{\partial x_2} }\right) \gamma_{R,R}(x_1,x_2,t) = 
\\
- \frac{i}{\sqrt 2} V_1 e^{ -i\Omega_{ge}t} \left[{ \delta (x_1) \beta_R(x_2,t) + \delta (x_2) \beta_R(x_1,t) }\right],
 \end{gathered}  \label{eq_gammaRR}
   \end{equation}
    \begin{equation}
 \begin{gathered}
\left({ \frac{d}{dt} - \upsilon_g \frac{\partial}{\partial x_1} - \upsilon_g \frac{\partial}{\partial x_2} }\right) \gamma_{L,L}(x_1,x_2,t) = 
\\
- \frac{i}{\sqrt 2} V_1 e^{ -i\Omega_{ge}t} \left[{ \delta (x_1) \beta_L(x_2,t) + \delta (x_2) \beta_L(x_1,t) }\right],
 \end{gathered}  \label{eq_gammaLL}
   \end{equation}
    \begin{equation}
 \begin{gathered}
\left({ \frac{d}{dt} + \upsilon_g \frac{\partial}{\partial x_1} - \upsilon_g \frac{\partial}{\partial x_2} }\right) \gamma_{R,L}(x_1,x_2,t) = 
\\
- \frac{i}{\sqrt 2} V_1 e^{ -i\Omega_{ge}t} \left[{ \delta (x_1) \beta_L(x_2,t) + \delta (x_2) \beta_R(x_1,t) }\right],
 \end{gathered}  \label{eq_gammaRL}
   \end{equation}            
where we have introduced the resonant frequency for the transition between the two upper levels, $\Omega_{ef}=\Omega_{gf}-\Omega_{ge}$. 

During the derivation we will assume initial conditions for a general-shape single-photon pulse $\beta_0(x)$ incident from the left side of the waveguide on an excited to the $\left| e \right\rangle$ state 3LE (see Fig. \ref{fig1}a):
   \begin{equation}
\begin{gathered}
\beta_R(x,0) = \beta_0(x), \qquad \beta_L(x,0)= \alpha(0)=0,
\\
\gamma_{i,j}(x_1,x_2,0)=0, \quad i,j=R,L.  \label{init}
\end{gathered}
   \end{equation}
Finding analytical expressions for all amplitudes allows us to provide a complete description of our model.

A detailed derivation for the similar problem of a spontaneously decaying 3LE, but in the case of a chiral waveguide, i.e., for photons moving only in the forward direction, was presented in \cite{Chuikin26}. Here, we describe the solution only briefly since we use the same mathematical approach, and it can be understood much more easily from \cite{Chuikin26}. Yet again, a similar problem of stimulated emission but for a regular two-level system and a Gaussian incident pulse was studied in \cite{Chuikin24}. The overall methodology is based on \cite{Mukhopadhyay24}, where the same approach was used to study multiphoton quantum Rabi oscillations.

One can easily find a formal solution for two-photon amplitudes from (\ref{eq_gammaRR}-\ref{eq_gammaRL}). For simplicity, we write here only the expression for $\gamma_{R,R}$: 
   \begin{equation}
\begin{gathered}
\gamma_{R,R}(x_1,x_2,t) = 
\\
-\frac{i}{\sqrt 2 } {V_1} \int\limits_0^t d\tau {e^{- i\Omega_{ge} \tau } } \delta (x_2 - \upsilon_g t') \beta_R(x_1 - \upsilon_g t',\tau )
\\
- \frac{i}{\sqrt 2 } V_1 \int\limits_0^t d\tau {e^{-i\Omega_{ge} \tau } } \delta (x_1 - \upsilon_g t') \beta_R(x_2 - \upsilon_g t',\tau ),
\end{gathered}
   \end{equation}
where $t' =t - \tau$. Two other amplitudes $\gamma_{L,L}(x_1,x_2,t)$ and $\gamma_{R,L}(x_1,x_2,t)$ can be found in a similar fashion.

Thus, the formal solution for the two-photon amplitudes (\ref{eq_gammaRR}-\ref{eq_gammaRL}) after time integration is as follows:
      \begin{widetext}

   \begin{subequations}
\begin{align}
\begin{gathered}
\gamma_{R,R}(x_1,x_2,t) = -i\frac{V_1}{\sqrt 2 \upsilon_g} e^{-i\Omega_{ge} \left({ t - \frac{x_1}{\upsilon_g} }\right)} \beta_R \left({ x_2 - x_1,t - \frac{x_1}{\upsilon_g} }\right) \theta (x_1) \theta \left( {t - \frac{x_1}{\upsilon_g}} \right)
 \\
 - i\frac{V_1}{\sqrt 2 \upsilon_g} e^{-i\Omega_{ge} \left({ t - \frac{x_2}{\upsilon_g} }\right) } \beta_R \left( {x_1 - x_2,t - \frac{x_2}{\upsilon_g}} \right)\theta (x_2) \theta \left({ t - \frac{x_2}{\upsilon_g} }\right),
\end{gathered}
\\
\begin{gathered}
\gamma_{L,L}(x_1,x_2,t) = -i\frac{V_1}{\sqrt 2 \upsilon_g} e^{-i\Omega_{ge} \left({ t + \frac{x_1}{\upsilon_g} }\right)} \beta_L \left({ x_2 - x_1,t + \frac{x_1}{\upsilon_g} }\right) \theta (-x_1) \theta \left( {t + \frac{x_1}{\upsilon_g}} \right)
 \\
 - i\frac{V_1}{\sqrt 2 \upsilon_g} e^{-i\Omega_{ge} \left({ t + \frac{x_2}{\upsilon_g} }\right) } \beta_L \left( {x_1 - x_2,t + \frac{x_2}{\upsilon_g}} \right)\theta (-x_2) \theta \left({ t + \frac{x_2}{\upsilon_g} }\right),
\end{gathered}
\\
\begin{gathered}
\gamma_{R,L}(x_1,x_2,t) = -i\frac{V_1}{\sqrt 2 \upsilon_g} e^{-i\Omega_{ge} \left({ t - \frac{x_1}{\upsilon_g} }\right)} \beta_L \left({ x_1 + x_2,t - \frac{x_1}{\upsilon_g} }\right) \theta (x_1) \theta \left( {t - \frac{x_1}{\upsilon_g}} \right)
 \\
 - i\frac{V_1}{\sqrt 2 \upsilon_g} e^{-i\Omega_{ge} \left({ t + \frac{x_2}{\upsilon_g} }\right) } \beta_R \left( {x_1 + x_2,t + \frac{x_2}{\upsilon_g}} \right)\theta (-x_2) \theta \left({ t + \frac{x_2}{\upsilon_g} }\right),
\end{gathered}
\end{align}  \label{gamma}
   \end{subequations} 
where the time integration limits give rise to Heaviside step functions, defined as $\theta(x>0) = 1,\quad \theta(x<0) = 0$, with the standard zero-point convention $\theta(0) = 1/2$. 

%We also remind that here we assume the initial amplitudes for two-photon states to be zero, $\gamma(x_1,x_2,0)=0$, since we consider only a single-photon incident pulse.

Then, substituting (\ref{gamma}) into (\ref{eq_betaR}), (\ref{eq_betaL}), we can derive expressions for $\beta_{R,L}(x,t)$. However, these expressions are not explicit because they contain time-retarded values of the same amplitude; for example, we obtain $\beta_R(x,t) \sim \beta_R(\upsilon_g t' -x,t-x/\upsilon_g)$. By performing the substitutions $x \to (\upsilon_g t' -x)$ and $t \to (t-x/\upsilon_g)$ and substituting these terms back into the original expression, one can obtain the final solution. This is possible because the recursion terminates due to the mismatch between the theta functions and the integration limits (see the appendix in \cite{Chuikin26} or \cite{Chuikin24}). The final analytical expressions for the single-photon amplitudes are:
   \begin{subequations}
\begin{align}
\begin{gathered}
\beta_R (x,t) = e^{ -\Gamma_1 t} \beta_0(x - \upsilon_g t) - i\frac{V_2}{\upsilon_g} e^{ - i\Omega_{ef} \left({ t - \frac{x}{\upsilon_g} }\right) } e^{-\Gamma_1 \frac{x}{\upsilon_g}} \theta (x) \theta \left({ t - \frac{x}{\upsilon_g} }\right) \alpha \left({ t - \frac{x}{\upsilon_g} }\right)
\\
- \Gamma_1 \theta (x) \theta \left({ t - \frac{x}{\upsilon_g} }\right) e^{-\Gamma_1 \left({ t - \frac{x}{\upsilon_g} }\right)} \int\limits_0^{x/\upsilon_g} {dt'} e^{ -\Gamma_1 t'} e^{- i\Omega_{ge} \left({ t' - \frac{x}{\upsilon_g} }\right)} \beta_0 \left[\upsilon_g(t' - t) \right],
\end{gathered}
\\
\begin{gathered}
\beta_L(x,t) =  -i\frac{V_2}{\upsilon_g} e^{-i\Omega_{ef} \left( t + \frac{x}{\upsilon_g} \right) } e^{\Gamma_1 \frac{x}{\upsilon_g}} \theta (-x) \theta \left({ t + \frac{x}{\upsilon_g} }\right) \alpha \left({ t + \frac{x}{\upsilon_g} }\right)
\\
 - {\Gamma _1}\theta ( - x)\theta \left({ t + \frac{x}{\upsilon_g} }\right) e^{-\Gamma_1 \left( {t + \frac{x}{\upsilon_g}} \right)} \int\limits_0^{-x/\upsilon_g} {dt'} {e^{ -{\Gamma _1}t'}} e^{ - i{\Omega_{ge}} \left({ t' + \frac{x}{\upsilon_g} }\right) } \beta_0 \left[ \upsilon_g(t' - t) \right],
\end{gathered}
\end{align} \label{beta}
   \end{subequations}

      \end{widetext}
where $\beta_0(x)=\beta_R(x,t=0)$ is the initial single-photon pulse, and we define the decay rates for the levels $\left| e \right\rangle$ and $\left| f \right\rangle$ as $\Gamma_{1,2}=V_{1,2}^2/\upsilon_g$. We also note that, due to the obvious symmetry, the left- and right-moving photon amplitudes differ only in the initial pulse and the coordinate transformation $x \rightarrow -x$. Thus, we can also write $\beta_R(x,t)$ as:
   \begin{equation}
\beta_R (x,t) = e^{ -\Gamma_1 t} \beta_0(x - \upsilon_g t) + \beta_L(-x,t).   
   \end{equation}
      
Finally, using expressions (\ref{beta}) in (\ref{eq_alpha}), we can obtain the last desired amplitude of our wavefunction, corresponding to a fully excited atom in the state $\left| f \right\rangle$:
   \begin{equation}
\alpha (t) = \alpha (0) e^{-\Gamma_2 t} - iV_2 e^{-\Gamma_2 t} \int\limits_0^t {d\tau} e^{\left({ i\Omega_{ef} - \Gamma_1 + \Gamma_2 }\right) \tau } \beta_0 (-\upsilon_g\tau).  \label{alpha}
   \end{equation}
Here, despite the (\ref{init}) we include the initial value $\alpha(0)$ for the sake of generality, but in all further calculations we take it to be zero. The case of spontaneous emission of two photons from a fully excited 3LE, i.e., for the initial conditions $\alpha(0) =1, \beta_0(x)=0$, was studied in \cite{Chuikin26} for a chiral waveguide.

\section{Scattering dynamic of the incident exponential pulse}

The obtained probability amplitudes are written for an arbitrary pulse shape $\beta_0(x)$. To study a more specific case, we assume that the incident photon is an exponential pulse. This type of pulse shape is very common in waveguide QED since photons with this shape are emitted in the process of spontaneous decay of atoms \cite{Valente12}. In real space, we can write it as follows:
   \begin{equation}
\begin{gathered}
\beta_0(x) = \sqrt{\Delta_k} e^{(ik_0+\Delta_k/2)x} \theta (-x) 
\\
= \left({ \frac{2}{\Delta_x^2} }\right)^{1/4} e^{ i\omega_0\frac{x}{\upsilon_g} } e^{\frac{x}{\sqrt 2 \Delta_x }}  \theta (-x), \label{beta0}
\end{gathered}
   \end{equation}
where $\omega_0 = k_0 \upsilon_g$ is the central frequency of the pulse, and $\Delta_k$ is the pulse width in momentum space. However, we will use the spatial width of the packet $\Delta_x = \sqrt 2/ \Delta_k$ or the spectral width $\Delta_\omega = \Delta_k \upsilon_g$ more often. Initial pulse shape is shown in Fig. \ref{fig3}a.

Substituting the exponential pulse (\ref{beta0}) into (\ref{alpha}), we obtain the amplitude of a fully excited atom in the state $\left| f \right\rangle$ with zero photons in the waveguide:
   \begin{equation}
\alpha (t) =  - i\sqrt{\Gamma_2 \Delta_\omega} \frac{ e^{i(\Omega_{ef} - \omega_0)t} e^{ - \left({ \Gamma_1 + \frac{\Delta_\omega}{2} }\right) t} - e^{ -\Gamma_2 t} }{ i(\Omega _{ef} - \omega_0) + \Gamma_2 - \Gamma_1 - \frac{\Delta_\omega}{2} },   \label{alpha_exp}
   \end{equation}
Note that if the frequency of the initial pulse $\omega_0$ is resonant with $\Omega_{ef}$, there is a special point at the pulse width $\Delta_\omega=2(\Gamma_2-\Gamma_1)$. In this case, both the numerator and denominator are equal to zero, so we need to evaluate the corresponding limit, which gives us $\lim_{\Delta_\omega \to 2(\Gamma_2-\Gamma_1)} \alpha(t)= -i\sqrt{2 \Gamma_2 (\Gamma_2 - \Gamma_1)} t e^{-\Gamma_2t}$.  
  
  \begin{figure}
\centering
\includegraphics[width=0.5\textwidth]{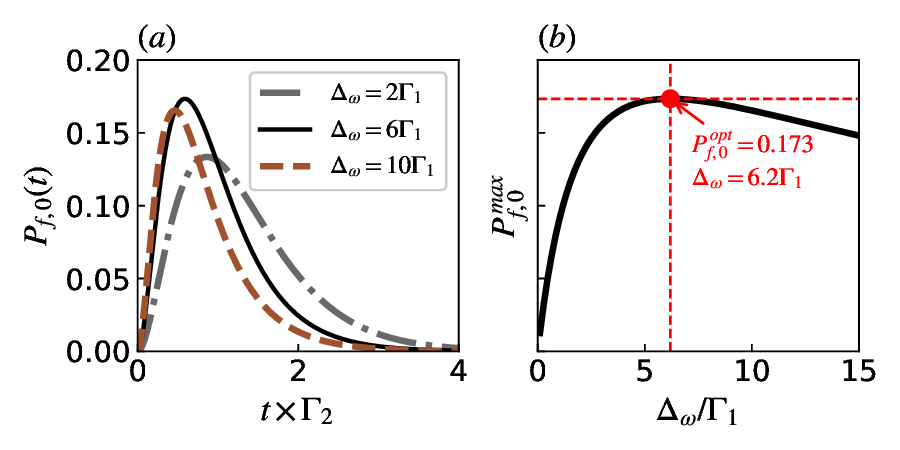}\\
  \caption{ (a) Probability of full excitation of 3LE by incident pulse for different spectral linewidth $\Delta_\omega$ plotted with expression (\ref{Pf0}). (b) Maximum probability of excitation versus impulse linewidth. For optimal linewidth $\Delta_\omega^{opt} = 6.179 \Gamma_1 \approx 6.2\Gamma_1$ we obtain maximum possible probability $P_{f,0}^{opt}=0.173$. Other parameters: $\Gamma_1/\Omega_{ge} = 0.02$, $\Gamma_2 = 1.5 \Gamma_1$, $\omega_0 = \Omega_{ef}$, $\alpha_r = -0.04$.}  \label{fig2} 
  \end{figure}

The probability of finding the atom in the state $\left| f \right\rangle$ with the waveguide in the vacuum state is:
   \begin{equation}
\begin{gathered}
P_{f,0}(t) = \left|{ \alpha (t) }\right|^2 = \frac{\Gamma_2\Delta_\omega}{ (\Omega_{ef} - \omega_0)^2 + \Gamma_-^2 } \times
\\
\left({  e^{ -(2\Gamma_1 + \Delta_\omega )t} + e^{-2\Gamma_2t} - 2e^{ -\Gamma_+ t} \cos (\Omega_{ef} - \omega_0)t } \right),
\end{gathered}  \label{Pf0}
   \end{equation}
where, for simplicity, we introduce $\Gamma_- = \Gamma_2 - \Gamma_1 - \Delta_\omega/2$, and $\Gamma_+ = \Gamma_2 + \Gamma_1 + \Delta_\omega/2$.

For further analysis we take typical parameters that are relevant to a superconducting transmon, $\Gamma_1/\Omega_{ge}=0.02$, $\Gamma_2=(1.5-2)\Gamma_1$ \cite{Koch07}. As we stated in the introduction section, nowadays transmon is the most popular type of artificial atom used in circuit QED and waveguide QED systems. It is convenient to describe its 3LE structure with a relative anharmonicity parameter $\alpha_r = (\Omega_{ge} - \Omega_{ef})/\Omega_{ge}$. For a typical transmon the anharmonicity is negative and very low, usually ranging from $-0.03$ to $-0.06$. However, later on we will consider more anharmonic systems also.

The probability of detecting the fully excited 3LE for different pulse widths is shown in Fig. \ref{fig2}a. The tallest peak is reached at $\Delta_\omega \approx 6 \Gamma_1$, whereas both narrower and broader pulses give a lower maximum, which indicates the existence of an optimal pulse duration. For Fig. \ref{fig2} we find the maximum obtained probability $P_{f,0}^{max}$ and plot it as a function of spectral linewidth $\Delta_\omega$. The curve exhibits a well-pronounced maximum at $\Delta_\omega^{opt} = 6.179 \Gamma_1$ that reaches $P_{f,0}^{opt} = 0.173$. For comparison, in the case of a two-level system excited by a Gaussian packet the maximum excitation probability reaches $P_{f,0}^{opt} = 0.4$ for an infinite waveguide and $P_{f,0}^{opt} = 0.72$ for a semi-infinite waveguide at a pulse width $\Delta_\omega = \Gamma$ \cite{Chen11}. Thus, for the three-level emitter the excitation probability is reduced due to the decay of the first-excited state $\left| e \right\rangle$.

Substituting the exponential pulse (\ref{beta0}) and using (\ref{alpha_exp}) in (\ref{beta}), we obtain explicit expressions for the single-photon amplitudes with the 3LE in the state $\left| e \right\rangle$:

\begin{widetext}

   \begin{equation}
\beta_R(x,t) = \left( \frac{2}{\Delta_x^2} \right)^{1/4} e^{-\Gamma_1t} {e^{- \left( {i\omega_0 + \frac{\Delta_\omega}{2} }\right) \left({t - \frac{x}{\upsilon_g} }\right) }}\theta \left( {t - \frac{x}{\upsilon_g}} \right) + {\beta _L}( - x,t),   \label{betaR_exp}
   \end{equation}  
   \begin{equation}
\begin{gathered}
\beta_L(x,t) = - \left( \frac{2}{\Delta_x^2} \right)^{1/4}  \frac{\Gamma_2 \theta (-x) \theta \left({ t + \frac{x}{\upsilon_g} }\right)}{ i(\Omega_{ef} - \omega_0) + \Gamma_2 - \Gamma_1 - \frac{\Delta_\omega}{2}  } \left({ e^{ -\Gamma_1 t} e^{ -\left({ i\omega_0 + \frac{\Delta_\omega}{2} }\right) \left({ t + \frac{x}{\upsilon_g} }\right)} - e^{\Gamma_1 \frac{x}{\upsilon_g} } e^{-\left({ i\Omega_{ef} + \Gamma_2 }\right) \left({ t + \frac{x}{\upsilon_g} }\right)} }\right)
\\
 - \left({ \frac{2}{\Delta_x^2} }\right)^{1/4} \frac{\Gamma_1 \theta (-x) \theta \left({ t + \frac{x}{\upsilon_g} }\right) }{ i(\omega_0 - \Omega_{ge}) + \frac{\Delta_\omega}{2} - \Gamma_1 } \left({ e^{- \Gamma_1 t} e^{- \left({ i\omega_0 + \frac{\Delta_\omega}{2} }\right)\left({ t + \frac{x}{\upsilon_g} }\right)} - e^{- \left({ i\omega_0 + \frac{\Delta_\omega}{2} + \Gamma_1 }\right) t} e^{- \left({ i\Omega_{ge} + \Gamma_1 }\right) \frac{x}{\upsilon_g} }  }\right).
\end{gathered}   \label{betaL_exp}
   \end{equation}

\end{widetext}

  \begin{figure}
\centering
\includegraphics[width=0.45\textwidth]{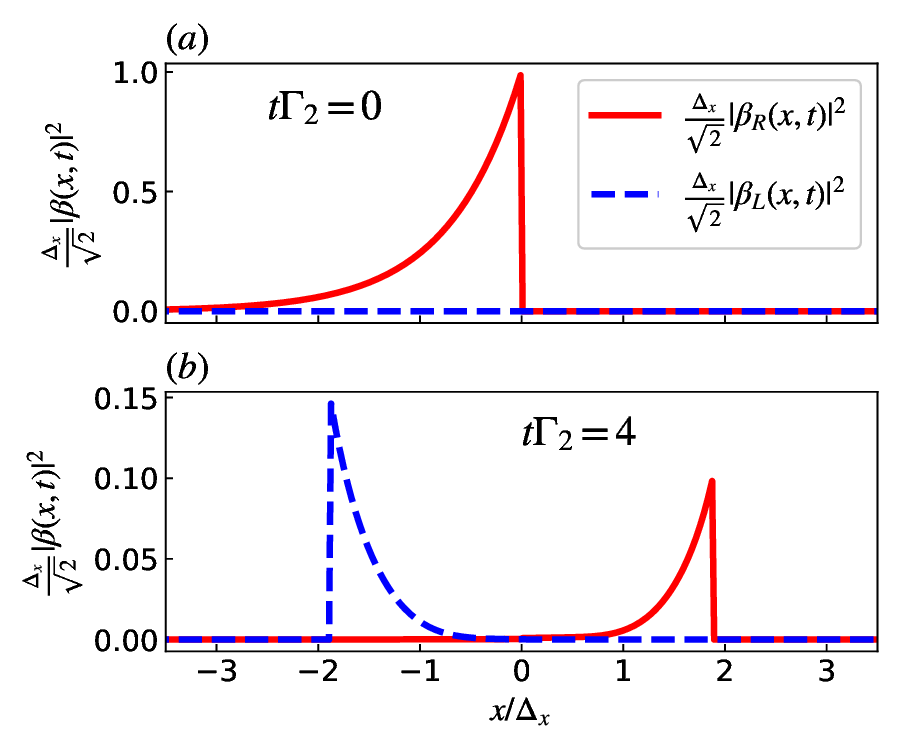}\\
  \caption{ Spatial distribution of single-photon states for a right-moving photon (red solid line) and a left-moving photon (blue dashed line) for (a) the initial moment $t = 0$ and (b) after scattering $t = 4/\Gamma_2$. The 3LE is located at $x=0$. Other parameters: $\Gamma_1/\Omega_{ge} = 0.02$, $\Gamma_2 = 1.5 \Gamma_1$, $\omega_0 = \Omega_{ge}$, $\alpha_r = -0.04$, $\Delta_\omega = \Gamma_1$. }  \label{fig3} 
  \end{figure}
  
The spatial distribution of single-photon states can be found as the square modulus of (\ref{betaR_exp}-\ref{betaL_exp}), and is plotted in Fig. \ref{fig3} for two moments in time. At the initial time $t=0$ the right-photon distribution is represented only by the initial pulse, $\beta_R(x,t)=\beta_0(x)$, and the reflected photon is absent, $\beta_L(x,t)=0$. After scattering both photons are represented by an exponential-shape pulses (Fig. \ref{fig3}b).

We can calculate the probabilities of finding a single photon (moving either to the left or to the right) and an atom excited to the state $\left| e \right\rangle$ by integrating the squared modulus of (\ref{betaR_exp}-\ref{betaL_exp}) over the entire waveguide:
   \begin{equation}
P_{e,R}(t) = \int\limits_{-\infty}^{+\infty} \left| \beta_R(x,t) \right|^2 dx, \quad P_{e,L}(t) = \int\limits_{-\infty}^{+\infty} \left| \beta_L(x,t) \right|^2 dx.  \label{Pe}
   \end{equation}

We can compare the scattering of a single photon from a 3LE with that from a regular qubit, or two-level system (TLS). In this way, one can identify the influence of the third level on the scattering dynamics. We can rewrite our results for an excited TLS by simply taking $\Gamma_2 = 0$, and consequently $\alpha(t)=0$. In this case, instead of (\ref{betaL_exp}), we obtain:
   \begin{equation}
\begin{gathered}
\beta_L^{TLS}(x,t) =  - \left({ \frac{2}{\Delta_x^2} }\right)^{1/4} e^{-\Gamma_1 t} \frac{\Gamma_1 \theta (-x) \theta \left({ t + \frac{x}{\upsilon_g} }\right) }{ i(\omega_0 - \Omega_{ge}) + \frac{\Delta_\omega}{2} - \Gamma_1 } \times
\\
 \left({ e^{- \left({ i\omega_0 + \frac{\Delta_\omega}{2} }\right)\left({ t + \frac{x}{\upsilon_g} }\right)} - e^{- \left({ i\omega_0 + \frac{\Delta_\omega}{2} }\right) t} e^{- \left({ i\Omega_{ge} + \Gamma_1 }\right) \frac{x}{\upsilon_g} }  }\right).
\end{gathered}   \label{betaL_tls}   
   \end{equation}   
The expression for (\ref{betaR_exp}) remains the same except for the last term, which should be replaced as $\beta_L(-x,t) \rightarrow \beta_L^{TLS}(-x,t)$. The same applies to the two-photon amplitudes (\ref{gamma}), where we simply replace $\beta_{R,L}(x,t)$ with $\beta_{R,L}^{TLS}(x,t)$.

  \begin{figure}
\centering
\includegraphics[width=0.5\textwidth]{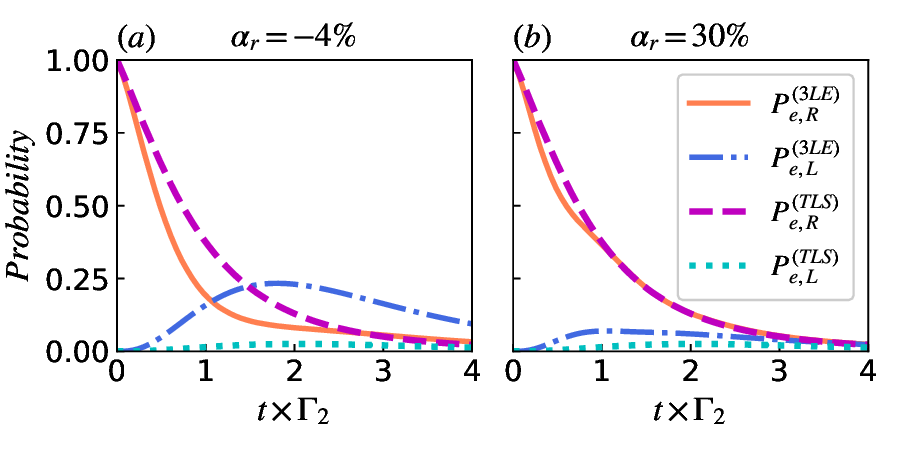}\\
  \caption{ Probability of finding a single photon in the waveguide (left or right) and the excited atom (3LE or TLS) in the state $\left| e \right\rangle$, calculated using the expressions (\ref{Pe}). (a) Small relative anharmonicity $\alpha_r = -0.04\%$, typical for a transmon; (b) larger anharmonicity $\alpha_r = 0.3\%$, usually associated with other types of artificial atoms. Both graphs are plotted for the optimal linewidth $\Delta_\omega = 6.2 \Gamma_1$ found in Fig. \ref{fig2}. Other parameters: $\Gamma_1/\Omega_{ge} = 0.02$, $\Gamma_2 = 3 \Gamma_1$, $\omega_0 = \Omega_{ge}$.}  \label{fig4} 
  \end{figure}
  
Comparison of the probabilities of finding single-photon states in the cases of scattering from a 3LE and a TLS is shown in Fig. \ref{fig4}. First, we consider a very low anharmonicity in our three-level system, where the resonant frequencies of the levels differ by only $4\%$ (Fig. \ref{fig4}a). This is a typical value for a transmon circuit, where the anharmonicity usually falls within the range from $-3\%$ to $-6\%$ \cite{Koch07}. In this case, we observe slightly different dynamical behavior between the 3LE and TLS. First, the influence of the level $\left| f \right\rangle$ enhances the decay of the excited state $\left| e \right\rangle$ by a small amount, which can be interpreted as some kind of superradiance induced not by another atom, but by an additional energy level. However, the effect is small, even for the optimal pulse width $\Delta_\omega = 6.2 \Gamma_1$ found earlier, but it can be increased by enlarging the $\Gamma_2/\Gamma_1$ ratio.

Secondly, the appearance of a left-moving photon is much more probable for a 3LE, as seen from the comparison of the blue dash-dotted and cyan dotted lines in Fig. \ref{fig4}a. This effect arises due to the possibility of reabsorption of the incident photon by the $\left| f \right\rangle$ level, which can then be emitted to the left via spontaneous decay. This behavior can be interpreted as coherent resonant scattering. It is well known that, in a 1D waveguide, an atom in the ground state acts as a nearly perfect mirror, fully reflecting an incident photon \cite{Greenberg15}, which was also proven experimentally \cite{Astafiev10a}. We observe a similar effect in our system between the levels $\left| e \right\rangle$ and $\left| f \right\rangle$, although it is largely suppressed by the decay of the middle level $\left| e \right\rangle$ with rate $\Gamma_1$. That said, the photon reflection is increased for larger $\Gamma_2/\Gamma_1$ ratios and can even surpass $P_{e,L}^{3LE}(t) = 0.5$ for $\Gamma_2/\Gamma_1 > 20$. Such large decay rates for the second-excited level are not typical for artificial atoms; however, they can be realized using a superconducting giant-atom architecture \cite{Vadiraj21}.

For larger anharmonicity $\alpha_r = 0.3$, the dynamics of the 3LE is almost identical to that of a TLS if the incident pulse is tuned to the lower transition, $\omega_0 = \Omega_{ge}$ (Fig. \ref{fig4}b). In this case, the second transition is far off resonance and practically does not interact with the incident photon. Albeit such a large anharmonicity is more relevant for other types of artificial atoms than transmons, e.g., superconducting flux qubits, fluxonium, or quantum dots.

\section{Suppression of stimulated emission by a third level}

Two-photon amplitudes can be found by substituting (\ref{betaR_exp}-\ref{betaL_exp}) into (\ref{gamma}). However, these explicit analytical expressions are too bulky and cumbersome, so we do not write them here and instead perform only a graphical analysis of the two-photon states.

  \begin{figure}
\centering
\includegraphics[width=0.48\textwidth]{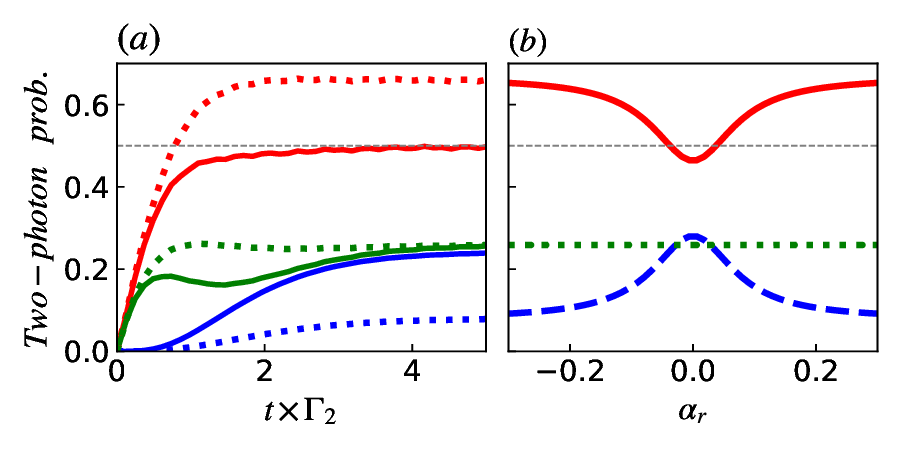}\\
  \caption{ (a) Probability of finding two right-moving photons $P_{RR}$ (red line), two left-moving photons $P_{LL}$ (blue line), or one right-moving and one left-moving photon $P_{RL}$ (green line). Solid lines describe scattering from a 3LE, while dotted lines correspond to scattering from a TLS. (b) Two-photon probabilities in the long-time limit $t\Gamma_2 = 15$ as a function of the 3LE anharmonicity parameter $\alpha_r$. The grey vertical line indicates a 50\% probability. Other parameters: $\Gamma_1/\Omega_{ge} = 0.02$, $\Gamma_2 = 1.5 \Gamma_1$, $\omega_0 = \Omega_{ge}$, $\Delta_\omega=3\Gamma_1$, for (a) $\alpha_r = -0.04$.}  \label{fig5} 
  \end{figure}

Similar to the case of single-photon states, the probability of finding two photons inside a waveguide is:
   \begin{equation}
P_{i,j}(t) = \iint\limits_{-\infty}^{+\infty} dx_1 dx_2 \left| \gamma_{i,j}(x_1,x_2,t) \right|^2, \qquad  i,j = R,L. \label{P_2phot}
   \end{equation}
We also note that, in the case of $P_{RL}$, we need to multiply the integral by 2 since we consider only the $\left| RL \right\rangle$ state. 

The time dependence of two-photon probabilities is shown in Fig. \ref{fig5}a. As expected, when $t = t_{\infty} \gg 1/\Gamma_2$, they approach constant values since atomic dynamics play a major role only at short times $t \sim 1/\Gamma_{1,2}$ after the scattering. In Fig. \ref{fig5}a, we also compare the probability of finding two photons in the case of scattering from a three-level system (solid lines) and from a regular qubit (dotted lines). 

We also want to state that the sum of all probabilities (\ref{Pf0}), (\ref{Pe}) and (\ref{P_2phot}) gives us unity for any arbitrary time with deviations less than 2\%. Thus, for our solution the norm of the wavefunction (\ref{wavefunc}) is preserved with sufficient accuracy.

The effect of stimulated emission is manifested in an increase of $P_{RR}$ over $P_{RL}$. From Fig. \ref{fig5}a, we observe an interesting effect: in the case of a low-anharmonicity three-level system, such as a superconducting transmon, stimulated emission is largely suppressed (compare the red solid and dotted lines). Indeed, for a two-level system, the maximum probability of finding both photons moving to the right (or in the same direction as the initial pulse) is $P_{RR}^{TLS}(t_{\infty}) \to 0.67$, see red dotted line in Fig. \ref{fig5}a. This result is in good agreement with \cite{Rephaeli12}, where a similar problem was considered in the stationary limit $t \to \infty$. However, if we add a third level, we see that the probability of finding two right-moving photons is much smaller, and, in fact, for $\alpha_r=-0.04$ it reaches 0.5. On the other hand, the probability of finding both photons moving in opposite directions, $P_{RL}(t_{\infty})$ (green line), remains the same regardless of whether the scattering occurs from a 3LE or a TLS.

In Fig. \ref{fig5}b, we analyse the long-time limit of the same probabilities as a function of the relative anharmonicity of the 3LE. The dependence is clearly symmetrical, so the sign of the anharmonicity parameter does not play a role here. As expected from the previous results for single-photon dynamics, larger anharmonicity leads to a closer resemblance between the 3LE and a qubit. We find that $\alpha_r = 0.3$ is sufficient to give practically the same values as those of a TLS (compare with the dotted lines in Fig. \ref{fig5}a). On the contrary, the biggest difference is observed for equidistant level spacing, i.e., when $\alpha_r = 0$. We also note that a state with two opposite-moving photons $\left| RL \right\rangle$ does not depend on anharmonicity at all (green dotted line in Fig. \ref{fig5}b).   

  \begin{figure}
\centering
\includegraphics[width=0.5\textwidth]{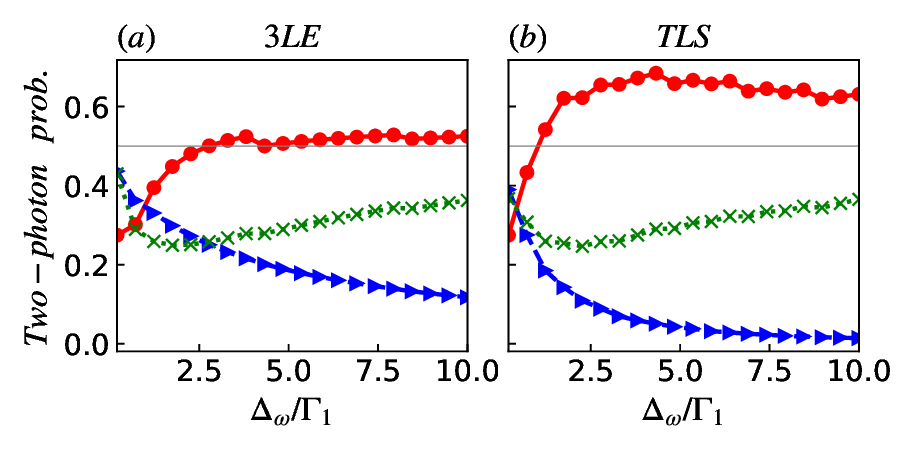}\\
  \caption{ Probability of finding two right-moving photons $P_{RR}$ (red solid line), two left-moving photons $P_{LL}$ (blue dashed line), or one right-moving and one left-moving photon $P_{RL}$ (green dotted line) for different incident pulse widths $\Delta_\omega$ in the case of scattering from a three-level system (a) or a regular two-level qubit (b). The time point $\overline{t_0} = t\Gamma_2$ is different for different regions, but is always chosen to be larger than 10. Other parameters: $\Gamma_1/\Omega_{ge} = 0.02$, $\Gamma_2 = 1.5 \Gamma_1$, $\alpha_r = -0.04$, $\omega_0 = \Omega_{ge}$.}  \label{fig6} 
  \end{figure}

We also study how the two-photon probabilities in the long-time limit depend on the spectral width of the incident pulse. These dependencies are shown in Fig. \ref{fig6} for both 3LE and TLS cases, although they are very similar in their properties. For a very short temporal duration of the pulse ($\Delta_\omega \to \infty$), $P_{LL} \to 0$, and $P_{RR} = R_{RL} \to 0.5$. Physically, this means that the incident pulse is so short in time that it does not interact with the atom at all, and we obtain two independent events: photon transmission through the atom and spontaneous atomic decay with equal probabilities in the left and right directions. For a very long pulse ($\Delta_\omega \to 0$), the photon interacts with an already decayed atom in the ground state, so it is fully reflected. In this case, $P_{RR} \to 0$, and $P_{LL} = R_{RL} \to 0.5$. Both 3LE and TLS atoms give similar results, which are in good agreement with the stationary solution \cite{Rephaeli12}.
  
For a very short pulse, when atomic decay and incident photon propagation can be viewed as separate events, the maximum probability of finding two right-moving photons is exactly 50\%. It is therefore reasonable to assume that any excess over $P_{RR} > 0.5$ can be interpreted as a stimulated emission effect. Our results in this section clearly show that the influence of the third level $\left| f \right\rangle$ suppresses the stimulated emission. From Fig. \ref{fig5}b, we can find the minimum required anharmonicity $\alpha_r > \alpha_r^{min}$ to obtain $P_{RR}>0.5$, and it is approximately $\alpha_r^{min} \approx \pm 0.04 =\pm 4\%$. This leads to the conclusion that superconducting transmons are not ideal candidates for tasks relying on the stimulated emission effect, and other types of artificial atoms, such as superconducting flux qubits or quantum dots, should be considered.

\section{Photon-photon correlations}

  \begin{figure*}
\centering
\includegraphics[width=0.8\textwidth]{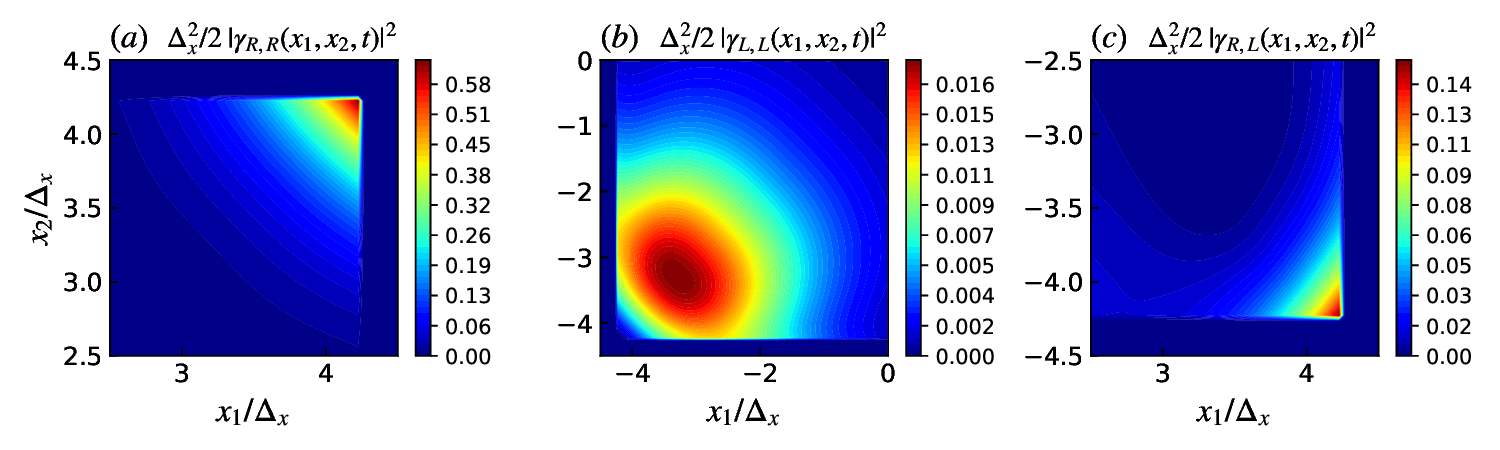}\\
\includegraphics[width=0.8\textwidth]{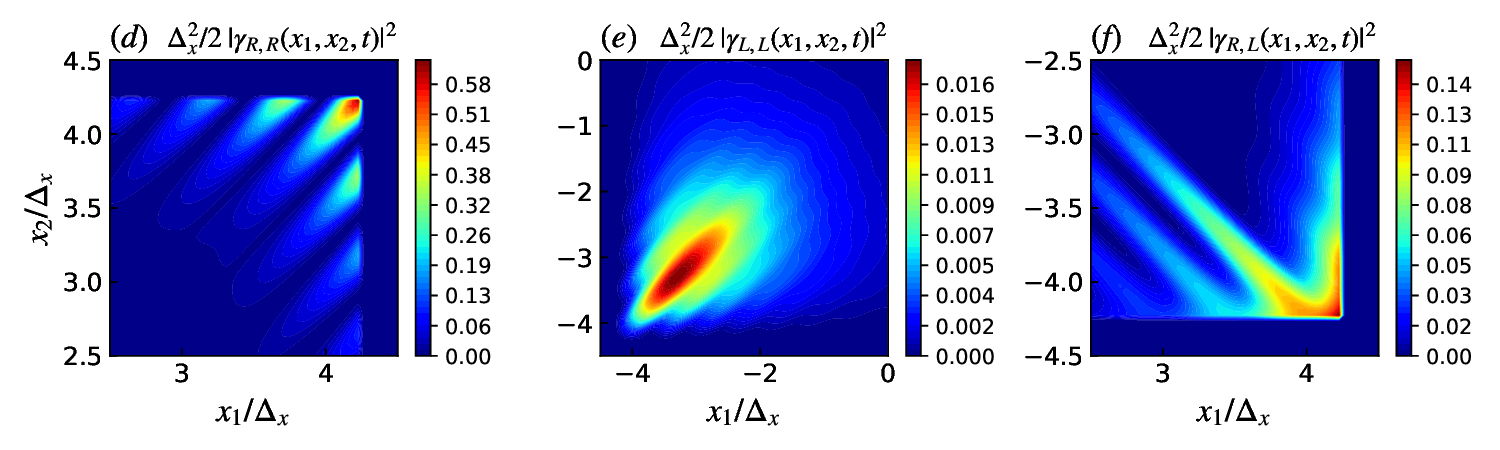}\\
  \caption{ Spatial distribution of two-photon states for: (a-c) a low-anharmonicity 3LE with $\alpha_r=-0.04$; (d-f) a high-anharmonicity 3LE with $\alpha_r = 0.5$, at a fixed time $t\Gamma_2 = 4$. Other parameters: $\Gamma_1/\Omega_{ge} = 0.02$, $\Gamma_2 = 2 \Gamma_1$, $\Delta_\omega = 3 \Gamma_1$, $\omega_0 = \Omega_{ef}$.}  \label{fig7} 
  \end{figure*} 
  
The other interesting effect that can be studied with our model is photon-photon correlation. The square modulus of the two-photon amplitudes (\ref{gamma}) gives the spatial distribution of the photonic fields inside the waveguide. The corresponding plots are shown in Fig. \ref{fig7} for both photons moving to the right (Fig. \ref{fig7}a,d), both photons moving to the left (Fig. \ref{fig7}b,e), or photons moving in opposite directions (Fig. \ref{fig7}c,f).

It is interesting that, for certain parameters, all three state distributions are localized at a single spatial point for $x_1=x_2$ (Fig. \ref{fig7}a-c). In the case of the $\left| RR \right\rangle$ and $\left| LL \right\rangle$ states, we can interpret this effect as photon bunching, meaning that both photons share not only the direction of propagation but also the same spatial coordinate. For the $\left| RL \right\rangle$ state, this indicates that both photons are located at the same distance from the origin $x=0$, but on opposite sides.

Photon bunching is obtained in the case of a low-anharmonicity 3LE, which we usually associate with a transmon artificial atom. For increased anharmonicity, more typical of other artificial atoms, we observe a spatial interference pattern formed in the density of the $\left| RR \right\rangle$ and $\left| RL \right\rangle$ states (Fig. \ref{fig7}d-f). Thus, complete bunching is reduced, except for the case when both photons move to the left, where the effect becomes even stronger. However, note that such a pronounced interference pattern is obtained for a very large anharmonicity $\alpha_r = 0.5$, which can be achieved only with a superconducting flux circuit.   

  \begin{figure}
\centering
\includegraphics[width=0.5\textwidth]{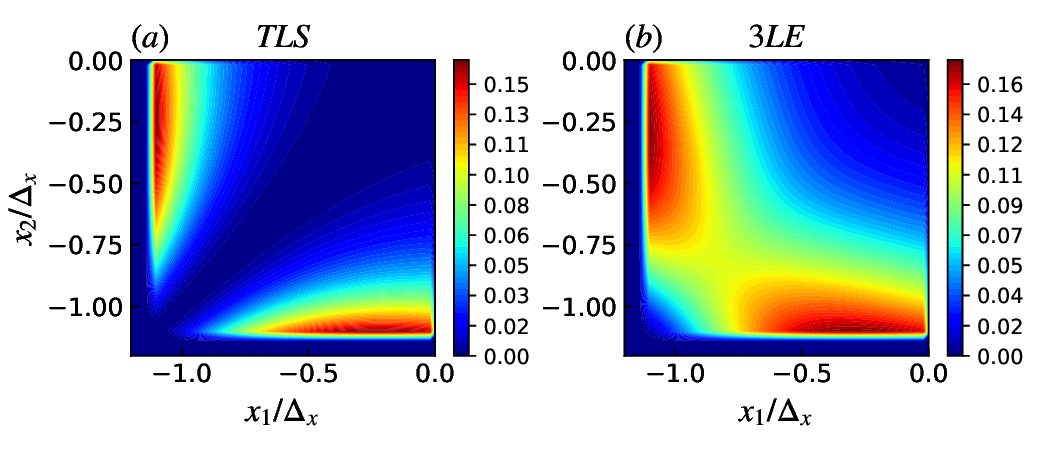}\\
  \caption{ Spatial distribution of a state with both photons moving to the left, $\Delta_x^2/2 |\gamma_{L,L}(x_1,x_2,t)|^2$, for a two-level system (a) and a three-level system (b). Other parameters: $t\Gamma_2 = 4$, $\Gamma_1/\Omega_{ge} = 0.02$, $\Gamma_2 = 2 \Gamma_1$, $\alpha_r = -0.04$, $\omega_0 = \Omega_{ge}$, $\Delta_\omega = 0.8 \Gamma_1$.}  \label{fig8} 
  \end{figure}
  
However, for a regular two-level system in a bidirectional waveguide, it is known that the transmitted field usually exhibits bunching, while the reflected field exhibits antibunching behavior \cite{Bradford14}. Indeed, if we replace the 3LE with a TLS, as we did before, the spatial distribution for two left-moving photons will have a zero dip along the diagonal line $x_1 = x_2$, see Fig. \ref{fig8}a. This is clearly a manifestation of the antibunching effect, meaning that two photons will occupy different spatial points in the waveguide. That said, this behavior is very different from that of a three-level system, for which we obtain the opposite effect of photon bunching (compare Fig. \ref{fig7}b with Fig. \ref{fig8}a).

However, the reflected two-photon field can demonstrate not only a bunching effect. In Fig. \ref{fig8}b, we plot the density of the $\left| LL \right\rangle$ state for a nearly monochromatic photon pulse resonant with $\Omega_{ge}$. For these parameters, we observe two distinct maxima at different spatial points, similar to the TLS case (Fig. \ref{fig8}a). Despite the fact that the diagonal dip does not fall completely to zero, we clearly observe a 'switching off' of the bunching behavior, as in Fig. \ref{fig7}b. Moreover, to achieve this, we only changed easily tunable parameters of the incident pulse, namely the central frequency $\omega_0$ and spectral width $\Delta_\omega$.

\section{Conclusion} 

In this work, we have studied the scattering of a single-photon pulse from a ladder-type three-level emitter in a one-dimensional open waveguide. Using the real-space formalism, we have derived analytical expressions for the system wavefunction and analysed the scattering dynamics for an exponentially shaped incident pulse. We have shown that the presence of the third energy level can strongly modify the stimulated emission process even for weakly anharmonic systems.

In particular, for parameters typical of superconducting transmon qubits, the third level significantly suppresses the generation of two co-propagating photons compared to the ideal two-level case. At the same time, the additional level leads to new two-photon correlation effects, including photon bunching in the reflected field. We have also shown that these correlation properties can be controlled by varying the incident photon parameters, such as the central frequency and spectral width. These results emphasize the importance of accounting for higher energy levels in artificial atoms and provide insight into two-photon scattering processes in waveguide quantum electrodynamics.

In this work we focused on scattering of exponential photon pulse. However, our results are obtained for general case and can be used to study interaction of a three-level or two-level systems with any type of a single-photon pulse, for example, incident Gaussian packet.

\begin{acknowledgments}

The authors thank A. A. Shtygashev and A. N. Sultanov for fruitful discussions and helpful comments on the manuscript. The work is supported by the Ministry of Science and Higher Education of Russian Federation under the project FSUN-2026-0004.

\end{acknowledgments}

%\appendix
%\section{Maybe some appendix will be needed}

\end{document}